\documentclass[11pt]{article}
\usepackage{amssymb}
\usepackage{amsmath}
\usepackage{amsthm}

\usepackage{graphicx,color}

\newcommand{\ket}[1]{\left | #1 \right \rangle}
\newcommand{\bra}[1]{\left \langle #1 \right |}

\def\openone{\leavevmode\hbox{\small1\kern-3.8pt\normalsize1}}

\def\cb{{\cal B}}

\theoremstyle{definition}

\newcommand{\proj}[1]{\ket{#1}\!\bra{#1}}

\newcommand{\poly}{{\rm poly}}

\begin{document}
\title{\LARGE\bf
IQP computations with intermediate measurements}

\author{Richard Jozsa$^1$, Soumik Ghosh$^2$ and  Sergii Strelchuk$^3$
%\footnote{Address from Sept 2024: Department of Computer Science, University of Oxford, Oxford OX1 3QG, U.K.}
\\[3mm]
  \small\it $^1$DAMTP, Centre for Mathematical Sciences, University of Cambridge,\\ \small\it Wilberforce Road, Cambridge CB3 0WA, U.K.\\
   \small\it $^2$Department of Computer Science, The University of Chicago,\\ \small\it Chicago, Illinois 60637, USA.\\
 \small\it $^3$Department of Computer Science, University of Oxford,\\ \small\it  Oxford OX1 3QG, U.K.\\
 [1mm]
 }

\date{}

\maketitle

\begin{abstract}
We consider the computational model of IQP circuits (in which all computational steps are $X$ basis diagonal gates), supplemented by intermediate $X$ or $Z$ basis measurements. We show that if we allow non-adaptive or adaptive $X$ basis measurements, or allow non-adaptive $Z$ basis measurements, then the computational power remains the same as that of the original IQP model; and with adaptive $Z$ basis measurements the model becomes quantum universal. Furthermore we show that the computational model having circuits of only $CZ$ gates and adaptive $X$ basis measurements, with input states that are tensor products of 1-qubit states from the set $\{ \ket{+}, \ket{1}, \frac{1}{\sqrt{2}}(\ket{0}+i\ket{1}), \frac{1}{\sqrt{2}}(\ket{0}+e^{i\pi/4}\ket{1})\} $, is quantum universal. In contrast to the relation of IQP to PH collapse, all our results here are manifestly stable under small additive implementational errors.

\end{abstract}

\section{Introduction and preliminary notations}\label{intro}
The IQP computational model was introduced by Shepherd and Bremner in \cite{BrSh} and it was subsequently shown \cite{BJS} that efficient classical simulation (up to multiplicative error) would imply collapse of the infinite tower of complexity classes known as the polynomial hierarchy, to its third level. In addition to its intrinsic theoretical interest, this result has inspired much further work on the model as a possible basis for demonstrating quantum computational advantage in near-term quantum computers \cite{IQP1,IQP2,IQP2.5,IQP3,IQP4,IQP5}.

The original IQP computational model is the following (cf \cite{BrSh,BJS}). We initialise $n$ qubits in state $\ket{0}^{\otimes n}$ and apply a sequence of $\poly (n)$ computational steps, each of which is a unitary gate diagonal in the $X$ basis. Finally we measure some, or generally all, qubits in the $Z$ basis to obtain the output. We refer to this formulation of the IQP model as the $X$-picture. We will not be fully prescriptive about the exact gates allowed but mention that for all results, it suffices to use only 1- and 2-qubit gates whose diagonal entries in the $X$ basis are all integer powers of $e^{i\pi/8}$ \cite{BJS}. 

There is an alternative equivalent formulation in terms of $Z$ basis diagonal gates, that we call the $Z$-picture:  the qubits are initialised in state $\ket{+}^{\otimes n}$ and $Z$ basis diagonal gates are applied before final output measurements in the $X$ basis. This $Z$-picture is readily obtained from the $X$-picture by inserting two `Hadamard bars' i.e.  two $H^{\otimes n}$ operations, at every stage in the $X$-picture, which then conjugate all $X$-diagonal gates into $Z$-diagonal gates, and modify the input state and output measurements as stated.

\begin{figure}[htb]
\begin{center}
\includegraphics[height=2.1in,width=6in,angle=0]{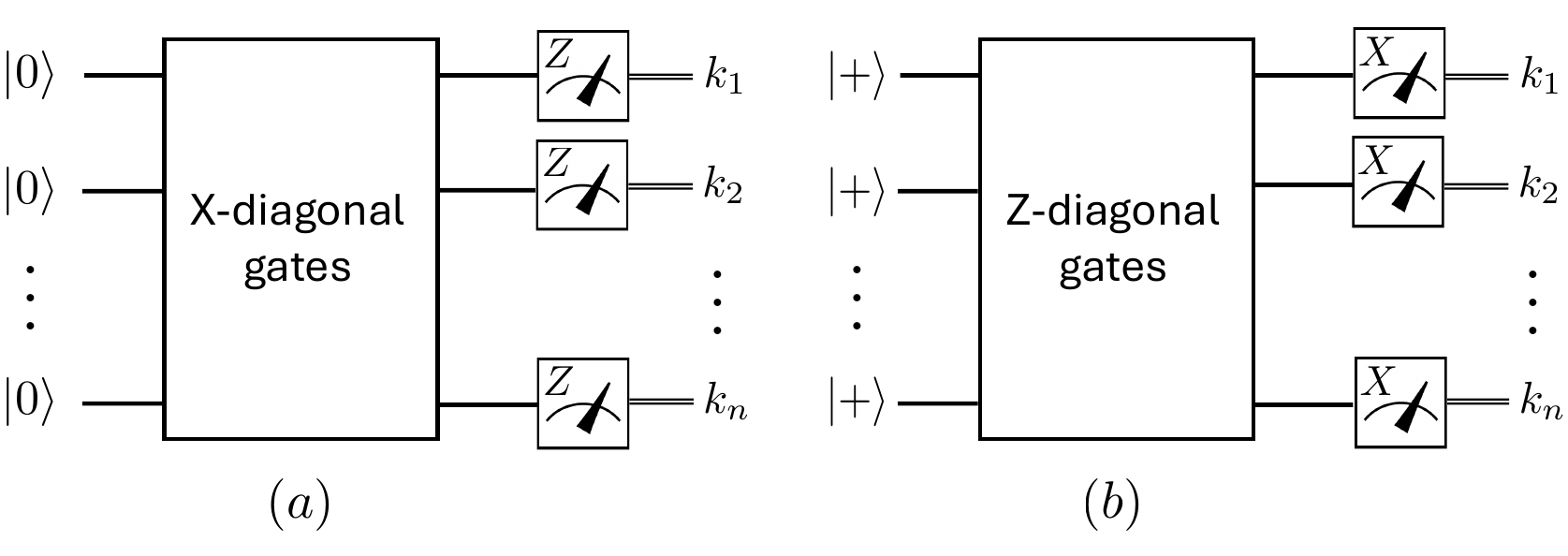}
\caption{The IQP computational model. (a) the $X$-picture, and  (b) the $Z$-picture. Double lines denote classical bit outputs of measurements.
}
\end{center}
\end{figure}

In this paper we will consider extensions of the IQP model in which $X$ basis or $Z$ basis measurements are additionally allowed as computational steps. These measurements can be {\em non-adaptive} or {\em adaptive}, the latter term referring to the further possibility that subsequent computational steps (unitary gates or measurements) can be chosen to depend on previous measurement outcomes. 

For $X$ and $Z$ measurements we associate integers $k=0,1$ to the outcomes via $k=0$ for $X$ outcome $\ket{+}$ and $Z$ outcome $\ket{0}$, and $k=1$ for $X$ outcome $\ket{-}$ and $Z$ outcome $\ket{1}$.

A notable characteristic  feature of the IQP model is that the computational steps all commute. In \cite{BrSh} this is referred to as the model being ``temporally unstructured''. In our extended models, this commutativity is broken in two ways: (i) $Z$ measurements do not commute with $X$-diagonal gates (although $X$ measurements do commute), and (ii) even though $X$ measurement operators commute with $X$-diagonal gates, a change of order cannot be implemented if the choice of $X$-diagonal gate is adaptive, depending on the $X$ measurement outcome. This feature of adaptivity imposing a time ordering on otherwise commuting actions, is also manifested in the PBC model of Bravyi, Smith and Smolin \cite{BSSPBCmodel} wherein all computational steps are pairwise commuting Pauli measurements but the sequence is adaptive.

The original IQP model has attracted much interest as a possible platform to demonstrate quantum advantage in near-term devices because of its connection to PH collapse (cf above). However this connection \cite{BJS}  is based on theoretical arguments involving post-selection, and as such, they require a representational accuracy of the circuit up to multiplicative (rather than additive) error, which presents a challenge for implementations. In view of this, we note that in contrast to the relation of IQP to PH collapse, our further results here do not rest on any arguments involving post-selection, and are thus manifestly stable under small additive implementational errors.

\section{Statement of main results and motivation}\label{mainresults}
{\bf Theorem 1.} Consider the IQP model (in the $X$-picture) extended to allow also non-adaptive or adaptive $X$ basis measurements, or $Z$ basis measurements, as computational steps in addition to $X$-diagonal unitary gates. Then these models have the following computational powers.\\
(a) non-adaptive or adaptive $X$ basis measurements: same power as the IQP model i.e. for each process of the extended type there is an IQP process (of the original type) that has the same output distribution;\\ (b) non-adaptive $Z$ basis measurements: same power as the IQP model;\\ (c) adaptive $Z$ basis measurements: universal quantum computing power.  $\Box$

We note that the original IQP model itself is not expected to have full universal quantum computing power, since if we measure only a single qubit (or $O(\log n)$ qubits) for the output, then the resulting process is classically efficiently simulatable \cite{BJS}.
With this in mind, the original motivation for our extended models here was the magic state and ``gate gadget'' formalism of Bravyi and Kitaev \cite{BrKit}. This provides a method for elevating the (sub-universal) computing power of Clifford computations to be fully quantum universal, in a way that does not enlarge the set of gates (Clifford gates) that are used. Instead, new exotic (non-stabiliser) input states (so-called magic states) are allowed, which, when used as inputs for Clifford circuits involving also intermediate adaptive measurements (so-called gate gadgets), allow the implementation of ``resourceful'' gates i.e. gates which, together with Clifford gates, form a universal set.

We can envisage applying these ideas to the (sub-universal) IQP model too. We will see in the proof of Theorem 1(c) that the already available $\ket{0}$ state (in the $X$-picture) can function as a magic state when used with a suitable gate gadget comprising  $X$-diagonal gates and an adaptive $Z$ measurement, to elevate IQP computing power to be fully quantum universal.

The magic state formalism is normally viewed as a means of elevating the power of a sub-universal quantum computational model without enlarging the gate set being used, but it can also be used to {\em reduce the size} of the gate set being used at the expense of expanding the variety of allowed input states (magic states). For example, in the scenario of Clifford gates we could use the same gate gadget as the $T$ gadget \cite{BrKit} but now with input magic state $\ket{0}+e^{i\pi/2}\ket{1}$, instead of $\ket{0}+e^{i\pi/4}\ket{1}$ that gives the $T$ gate, (c.f. Figure 3 below with $\alpha = \pi/4$ and $\pi/2$), and with adaptive correction now being the $Z$ gate (instead of $S$ that arises for the $T$ gate), to implement the $S$ gate wherever it is needed, thereby eliminating the need to include $S$ in the gate set being used. 

The apotheosis of this procedure is perhaps akin to the idea of measurement based quantum computing, in which all unitary gates have been eliminated and every computational step is a (generally adaptive) choice of measurement. However we can stop the reduction procedure before eliminating all unitary gates, and aim to retain a set of unitary gates, and set of magic states, that are both suitably ``simple'' in some useful sense. Applying these ideas in the setting of the IQP model we will prove the following result.

\noindent {\bf Theorem 2.}  Consider the computational model in which all computational steps are either $CZ$ gates or 1-qubit $X$ basis measurements (used both as generally adaptive intermediate measurements and for the output measurements), and all input states are tensor products of the four 1-qubit states  $ \ket{+}, \ket{1}, \frac{1}{\sqrt{2}}(\ket{0}+i\ket{1}), \frac{1}{\sqrt{2}}(\ket{0}+e^{i\pi/4}\ket{1}) $.\\ Then this model has universal quantum computing power.  $\Box$

Note that this model is a special case of the extended IQP model  of the kind that occurs in Theorem 1(c) (but here in the $Z$-picture, so the $Z$ measurements and $X$-diagonal gates in Theorem 1(c) now become $X$ measurements and $Z$-diagonal gates). Here, the set of diagonal gates has been reduced to allowing only the $Z$-diagonal $CZ$ gate but new kinds of input states (magic states) beyond those of the original IQP model are allowed now too.

In the following sections we give proofs of these Theorems and also develop some further properties of the IQP model along the way.

\section{The IQP model with intermediate measurements}\label{intmmts}

\subsection{Non-adaptive and adaptive $X$ measurements}
To aid readability we will sometimes use two notations for the $X$ basis states, writing $\ket{+}$ as $\ket{p}$ and $\ket{-}$ as $\ket{m}$. Unless otherwise stated, we always use the $X$-picture for the IQP model.

We begin by showing that the inclusion of adaptation for $X$ measurements (in the $X$-picture) does not bring any extra computational power over non-adaptive measurements. Consider the space $A$ of $n$ qubits with $X$ basis $\{ \ket{i_1 \ldots i_n}: i_1, \ldots ,i_n = p \mbox{ or } m \}$. For any fixed choice of $k=1, \ldots ,n$ we have the orthogonal decomposition $A=A_p\oplus A_m$ where $A_p$ (resp. $A_m$) is the span of all $\{ \ket{i_1 \ldots i_n}$ having $i_k=p$ (resp. $m$). Now, any $n$-qubit $X$-diagonal unitary $U$ preserves these subspaces, acting as $U=U_p\oplus U_m$ where $U_p$ (resp. $U_m$) is defined by all diagonal entries $e^{i\theta_{i_1\ldots i_n}}$ of $U$ having $i_k=p$ (resp. $i_k=m$). In terms of this decomposition we have the following result.

\noindent{\bf Lemma 1.} (Elimination of $X$ measurement adaptation.)\\
Consider the following adaptive process on any $n$-qubit state.\\
Process A:  do an $X$ measurement on line $k$ obtaining the outcome $x= p$ or $m$. Then, on the $n$-qubit post-measurement state, apply the $n$-qubit $X$-diagonal unitary $V(p)$ if $x=p$ or $V(m)$ if $x=m$. End of process.\\
This adaptive process gives the same result (i.e. same final state with same probability) as the following non-adaptive process.\\
Process NA:  do an $X$ measurement on line $k$ (and we do not need to note the outcome). Then, on the $n$-qubit post-measurement state, apply the $n$-qubit $X$-diagonal unitary $W$ defined by $W=V(p)_p\oplus V(m)_m$. End of process.

\noindent {\bf Proof.}  In both cases the initial $X$ measurement fixes the $k^{\rm th}$ line to be $\ket{x}$ for measurement outcome  $x=p$ or $m$, which is preserved by any subsequent $X$-diagonal unitary gate. Then the result follows directly from the definitions of $V(p)_p$ and $V(m)_m$ as the actions of $V(p)$ on $A_p$ and $V(m)$ on $A_m$. $\Box$

Hence any IQP circuit with adaptive $X$ measurements can be reduced to an equivalent IQP circuit with only non-adaptive measurements.

$X$ measurements commute with $X$-diagonal gates, so in an IQP circuit we can move them all to the start where they act on $\ket{0}$ states. Then they can be replaced by $Z$ measurements, as follows, which will be used in the proof of Theorem 1(a) below.

\noindent{\bf Lemma 2.}  (Replacing initial $X$ measurements by $Z$ measurements).\\
Let $\widetilde{CZ}$ denote the $X$-diagonal gate with matrix diag(1 1 1 -1) in the $X$ basis. Then 
\[ \widetilde{CZ}\ket{0}\ket{0}=\frac{1}{\sqrt{2}} \left( \ket{0}\ket{p}+\ket{1}\ket{m}\right). \]
Moreover for the latter state, a $Z$ measurement on the first qubit is exactly equivalent to an $X$ measurement on the second qubit (both having the same post-measurement states and probabilities).

\noindent {\bf Proof.} Immediate by inspection. $\Box$

\noindent {\bf Proof of Theorem 1(a).}\\
Suppose we have an IQP circuit $C$ of $X$-diagonal gates with (possibly adaptive) $X$ measurements too. The input state is $\ket{0}^{\otimes n}$ and the output is obtained by final $Z$ measurements.

Using Lemma 1 we rewrite the circuit as an equivalent IQP circuit with only non-adaptive measurements. Then we move all these measurements to the start and using Lemma 2, we replace them by  $Z$ measurements on new ancillary lines, each initially in state $\ket{0}$. All of these $Z$ measurements can then be moved to the end of the circuit as no further gates are applied to the new ancillary lines after the $\widetilde{CZ}$ gates.

This results in a slightly enlarged IQP circuit $C'$, now of the standard type (i.e. having no intermediate measurements), whose output (on the original non-ancillary lines) is equivalent to that of $C$. Furthermore the translation of the description of $C$ to that of $C'$ can clearly be done in polynomial time in the size of the circuit $C$. $\Box$

\subsection{Non-adaptive  $Z$ measurements}
We begin by establishing a more general IQP extension result of which the non-adaptive $Z$ measurement case will be a simple corollary. A {\em generalised permutation} $A$ of  a basis $\cb_K = \{ \ket{k}: k\in K \}$ is defined to be a permutation up to overall phases i.e. a mapping of the form $A:\ket{k} \mapsto e^{i\theta_k}\ket{P(k)}$ where $P$ is a permutation of $K$. Note that $A$ is generally not a $\cb_K$-diagonal gate but we always have $A=U\tilde P$ (or ${\tilde P}V$) where $\tilde P$ is a basis permutation and $U$ (and $V$) are $\cb_K$-diagonal.

\noindent {\bf Lemma 3.} Let $A$ on $d$ qubits be any generalised permutation of the $d$-qubit $X$ basis. Then the computational power of IQP (in the $X$ picture) extended to allow arbitrary use of $A$ as a computational step, is the same as that of IQP.

\noindent {\bf Proof.}  Note that these generalised permutations preserve $X$-diagonal-ness of gates under conjugation i.e. $\tilde{U}= AUA^\dagger$ is $X$-diagonal if $U$ is (and the diagonal entries of $\tilde{U}$ can be easily determined). Then using 
\begin{equation}\label{conjugate}
A\,U=(AUA^\dagger)\,A
\end{equation}
we can commute all uses of $A$ in an IQP circuit out to the left, updating each $X$-diagonal gate encountered as we go, leaving $A$ as the first gate acting on input $\ket{0}^{\otimes d}$. However $A=V\tilde{P}$ for $X$-diagonal $V$ and permutation $\tilde{P}$, and  $\tilde{P}$ clearly preserves 
\[ \ket{0}^{\otimes d}=  \left(\frac{1}{\sqrt{2}}(\ket{p}+\ket{m})  \right)^d =          \frac{1}{\sqrt{2^d}}\sum_{i_1, \ldots  i_d\, =\, p,\, m} \ket{i_1\ldots i_d} \]
 so we can replace $A$ by just its $X$-diagonal  part $V$. The resulting circuit then contains only $X$-diagonal gates as computational steps, giving the result. $\Box$
 
 \noindent {\bf Corollary (Theorem 1(b)).} The IQP model extended with intermediate $Z$ measurements that are non-adaptive, still has only IQP computing power.
 
 \noindent {\bf Proof.} For each $Z$ measurement, on line $k$ say, we introduce a new ancilla line $\ket{0}_a$ and replace the measurement operation by $CX_{ka}$. Thus we obtain a unitary circuit of $CX$ and $X$-diagonal gates. $CX$ is not $X$-diagonal but it does act as a permutation of the $X$ basis states of two qubits. Then Lemma 3 gives the result (with $V$ in the proof there being trivial). $\Box$
 
 In view of the proof of Lemma 3 it is interesting to consider which gates can be freely moved through $X$-diagonal gates via the adapting conjugation action in eq. (\ref{conjugate}). We have:
 
 \noindent{\bf Lemma 4.}  A unitary gate $W$ preserves the set of $X$-diagonal gates under conjugation iff $W$ is a generalised permutation of the $X$ basis.\\ Stated otherwise, the normaliser of the subgroup of $X$-diagonal gates in $U(2^n)$ is the group of generalised permutations of the $X$ basis. 
 
 \noindent {\bf Proof.}  The reverse implication is clear. For the forward implication,
 we work with matrices (of size $2^n \times 2^n$) which are matrices of the operators relative to the $X$ basis.
 Thus suppose $A$ is unitary and
\begin{equation}\label{diag} D’= A D A^\dagger \end{equation}
has $D’$ diagonal for any $D$ unitary diagonal. We will show that $A$ must then be a generalised permutation.

Since conjugation always preserves the set of eigenvalues of a matrix, it follows that the diagonal entries of $D’$ must be a permutation of those of $D$.
Now eq. (\ref{diag}) gives a linear map $L$ from the diagonal of $D$ to the diagonal of $D’ $ (each viewed as $2^n$--length vectors of complex numbers), and amongst all unitary $D$’s, there is an orthonormal basis of these diagonal vectors (e.g. take the diagonals to be the rows of the $n$-qubit $QFT$ matrix). Hence we can (by linearity) extend the conjugation action $L$ in eq. (\ref{diag}) from unitary diagonal $D$’s to all diagonal $D$’s (not just unitary ones).

Then choose $D$ to have a single 1 in the $k^{\rm th}$ diagonal place and 0’s elsewhere. So $D$ is the projector onto the corresponding X basis state $\ket{e_k}$, and the diagonal matrix $D’$ is the projector onto $A\ket{e_k}$. But the latter also has a single 1 on the diagonal, so is projector onto another $X$ basis state $\ket{e_m}$ for some $m$. Hence $A|\ket{e_k}$ must equal   $\ket{e_m}$ up to a phase (since $\proj{u}=\proj{v}$ iff $\ket{u}$ and $\ket{v}$ differ by an overall phase). Similarly for all choices of $k$, and we see that $A$ must be a generalised permutation matrix.  $\Box$

\subsection{Adaptive $Z$ measurements and the Hadamard gadget}
Having now proven Theorem 1(a) and (b), we turn here to Theorem 1(c).

\noindent{\bf Lemma 5.}  The IQP model extended to allow the use of the Hadamard gate $H$ as a computational step, is universal for quantum computation.

\noindent {\bf Proof.} The Hadamard operation has the same matrix in both the $X$ and $Z$ bases (since $H\ket{p}=\frac{1}{\sqrt{2}}(\ket{p}+\ket{m})$ etc.)  By a standard result on Clifford operations (c.f. \cite{NC} Section 4.5.3) it is known that the set of matrices \{matrix of $H$, diag(1, $e^{i\pi/4}$), diag(1 1 1 -1)\} is universal for all matrices in $U(2^n)$. Then the Lemma follows by interpreting these matrices as matrices of operations in either the $X$ or $Z$ basis.  $\Box$

In the IQP model we will implement $H$ using the so-called Hadamard gadget of \cite{BJS}, shown in Figure 2 for both the $X$- and $Z$-pictures. Let $\widetilde{CZ}$ denote the $X$-diagonal gate with matrix diag(1 1 1 -1) in the $X$ basis. The Hadamard gadget in the $X$-picture involves an application of $\widetilde{CZ}$, then a $Z$-measurement followed by a correction operator $Z$ applied adaptively if the $Z$-measurement outcome was 1. Its correctness can be easily verified by direct calculation. Note the the correction operator ($Z$ in the $X$-picture or $X$ in the $Z$-picture) is not available within the IQP model and its application will be treated by other means (c.f. Lemma 6 below).

In the Hadamard gadget given in \cite{BJS}, the correction operator does not appear, as there, the measurement output is {\em post-selected} to value 0. In our present work we do not include post-selection and the correction operator is then needed to deal with the ``wrong'' measurement outcome.

\begin{figure}[htb]
\begin{center}
\includegraphics[height=2in,width=6in,angle=0]{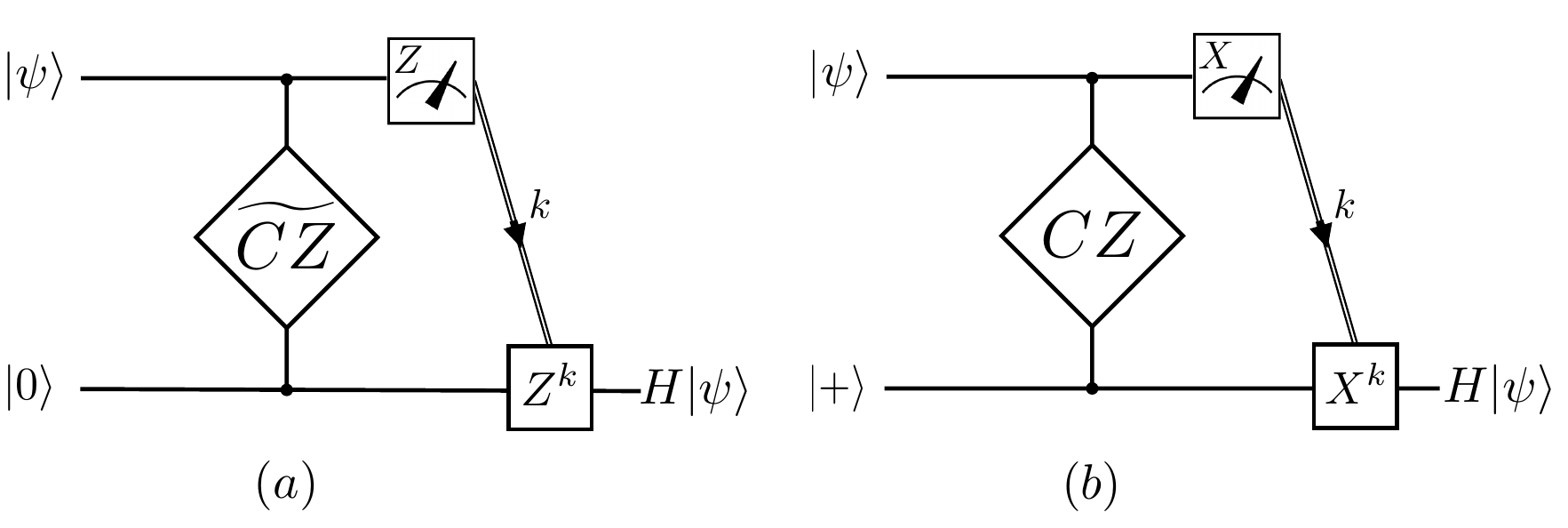}
\caption{The Hadamard gadget for (a) the $X$-picture, and (b) the $Z$-picture. The diagonal 2-qubit gates ($\widetilde{CZ}$ in the $X$-picture and $CZ$ in the $Z$-picture) are symmetric with regard to control and target qubit lines and are depicted with symmetrical symbols. Double lines represent classical bits (the measurement outcomes) which adaptively control the correction operators.
}
\end{center}
\end{figure}

\noindent {\bf Lemma 6 (Theorem 1(c)). } The IQP model (in the $X$-picture) extended to allow use of adaptive $Z$-measurements as computational steps, is universal for quantum computation.

\noindent {\bf Proof.} By Lemma 5, for universality it suffices to have $H$ available with $X$-diagonal gates. We implement $H$ using the Hadamard gadget which, in addition to an $X$-diagonal gate, involves an adaptive $Z$ measurement and a possible $Z$ correction operation. The latter is not available in the computational model as an allowed gate so we proceed as follows.

 $Z$ acts as a permutation on the $X$ basis. Thus we can use the conjugation procedure in the proof of Lemma 3 to commute the $Z$ gate out to the right (rather than to the left as in Lemma 3's proof), adapting each $X$-diagonal gate as we proceed (and leaving any subsequent $Z$ measurement steps unchanged). Then with $Z$ finally placed immediately before the final output $Z$ measurements, we simply delete it, as it has no effect on those measurement outcomes.

We apply this procedure successively for each occurrence of $H$, which finally leaves a circuit of only $X$-diagonal gates and adaptive $Z$ measurements, that can implement universal quantum computation. $\Box$

\section{Computational universality of circuits having only $CZ$ gates and adaptive $X$ basis measurements}\label{CZadXmmt}
In this section we give a proof of Theorem 2. Introduce the notations
\[ \ket{\alpha} = \frac{1}{\sqrt{2}}(\ket{0}+e^{i\alpha}\ket{1}) \hspace{2mm}\mbox{and} \hspace{2mm}
P_\alpha = \left( \begin{array}{cc}
1 & 0 \\ 0 & e^{i\alpha}
\end{array}
\right) \hspace{2mm} \mbox{(in the $Z$ basis)}.
\]
The statement of Theorem 2 relates to the IQP model in the $Z$-picture. We will use the Hadamard gadget (in its $Z$-picture version) as well as a further gadget called the $P_{\alpha}$ gadget, which is shown in Figure 3.  Its claimed action can be easily verified directly. Note that the required input ``magic state'' can be made from $\ket{\alpha}$ by applying a Hadamard gadget to it.

\begin{figure}[htb]
\begin{center}
\includegraphics[scale=0.5,angle=0]{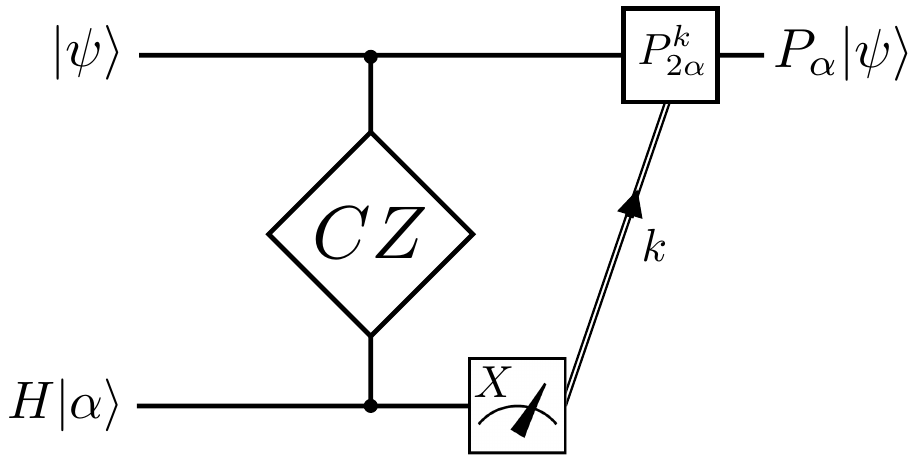}
\caption{The $P_\alpha$ gadget. It implements the phase gate $P_\alpha$ in terms of a $CZ$ operation, an adaptive $X$ measurement and a $(2\alpha)$-phase gate correction, while also consuming the input ``magic state'' $H\ket{\alpha}$.
}
\end{center}
\end{figure}

Now, consider any circuit of $H$, $T$ and $CZ$ gates (a scenario which provides universal quantum computation).

We begin by replacing each $H$ gate by a Hadamard gadget and each  $T$ gate by a $T=P_{\pi/4}$  gadget. This will require $\ket{+}$ and $\ket{\pi/4}$ state inputs, and and possible $X$  (from the $H$ gadget)  and $S=T^2$ (from the phase gadget) correction gates, as well as  $CZ$'s and adaptive $X$ measurements (that are allowed in Theorem 2).

Next, similarly, we replace each $S$ gate by an $S=P_{\pi/2}$ gadget. This will require $\ket{\pi/2}$ state inputs, possible adaptive $Z=S^2$  and $X$ gate correction operations, and further $CZ$'s and adaptive $X$ measurements.

The resulting circuit then comprises only $CZ$ gates, adaptive $X$ measurements and inputs $\ket{+}$, $\ket{\pi/2}$ and $\ket{\pi/4}$ (all allowed in Theorem 2), as well as $X$ and $Z$ gates.

To eliminate the $X$ gates, we commute them out to the right, using $ZX=-XZ$ (giving just an irrelevant overall phase) and 
$(CZ_{ab})(X_a)=(X_aZ_b)(CZ_{ab})$ (and similarly with labels $ab$ on the $CZ$'s reversed). Then when $X$ has been moved to immediately precede an $X$ measurement we just delete it, as it has no effect there. Finally to eliminate the $Z$ gates, we introduce a (single copy of) input state $\ket{1}$ and replace each $Z$ by a $CZ$ gate.
The resulting circuit then comprises only $CZ$ gates, (possibly adaptive) $X$ measurements and inputs from the given list, as required. $\Box$

Finally we make some remarks \cite{YukiComments} on the resource ingredients that appear in the statement of Theorem 2. 

The input state $\ket{+}$ is consumed in  $H$ gadgets and in phase gadgets (in their magic state preparations), so generally polynomially many input $\ket{+}$ states will be required. But only a {\em single} copy of the state $\ket{1}$ is required (cf above proof).
We may attempt to remove the need for $\ket{1}$ altogether e.g. using $\ket{1}=XH\ket{+}$ thereby obtaining it from another use of $\ket{+}$. But this will not work -- the needed subsequent commutation of $X$ gates in that construction, out to the right (across $CZ$'s), will introduce further new $Z$ gates that will not yet have been re-expressed as $CZ$'s, requiring availability of a further new $|1>$ state, etc.  {\em ad infinitum}.

Figure 3(b) of \cite{YT} gives an alternative gate gadget (comprising only $H$ and $CZ$ gates, with the latter shown as $CCZ$ with the first control set to $\ket{1}$) for implementing $S$. As for our $S$ gate gadget, it also uses a $\frac{1}{\sqrt{2}}\left( \ket{0}+i\ket{1}\right)$ state, but unlike our gadget it uses it {\em catalytically}. Other possible gate gadgets of this type for implementing the $S$ gate have been given in \cite{GF,AGP}.
Hence, like the state $\ket{1}$  in the list of input states in Theorem 2, it suffices to have only a {\em single} copy of $\frac{1}{\sqrt{2}}\left( \ket{0}+i\ket{1}\right)$ available too.

\subsection{Relation to measurement based quantum computing models}
The computational model in Theorem 2 can be interestingly related to some standard measurement based models of quantum computing.

Consider first the seminal measurement based model of Raussendorf and Briegel \cite{RB} based on the so-called cluster state. This model may be summarised as follows. (a) We begin with the multi-qubit cluster state which is the state obtained by applying $CZ$ gates to a suitable array of $\ket{+}$ states. (b) Then to achieve universal quantum computation, these $CZ$ gates can be leveraged to implement $CZ$ gates of a quantum computation, and further 1-qubit gates (such as $T$ and $H$) are implemented by patterns of adaptive 1-qubit measurements (from a restricted set), applied to cluster state qubits.

The parallel with Theorem 2 is then evident, involving a shift of input states allowed and set of measurements performed: the measurement patterns for 1-qubit gates can be viewed as gate gadgets with the cluster state serving as input ``magic state'', and the cluster state itself viewed as the result of a unitary circuit of $CZ$ gates, acting on inputs restricted now to be only $\ket{+}$ states. Indeed our $H$ gadget (in the $Z$-picture, with its $\ket{+}$ state ``magic'' input) is then essentially the same as the measurement pattern for $H$ in the Raussendorf-Briegel model.

As a second measurement based model, recall the Pauli based computing model of Bravyi, Smith and Smolin \cite{BSSPBCmodel} (and c.f. also \cite{YJS} for an extended exposition). In this model the input state has the form $\ket{\pi/4}^{\otimes n}$ and universal computing is achieved by an adaptive sequence of $n$-qubit Pauli measurements that all pairwise commute. The starting point for establishing  this model is the Bravyi-Kitaev implementation of universal computation in terms of Clifford circuits with $\ket{\pi/4}$ magic state inputs (giving $T$ gates via $T$ gadgets), and then commuting all unitary Clifford gates out to the right, to after the final output measurements (where they can be deleted). In this process the 1-qubit adaptive Pauli measurements in the $T$ gadgets are conjugated into generally $n$-qubit Pauli measurements which, with further processing \cite{BSSPBCmodel,YJS} can be cut down to leave only mutually commuting ones.

Now note that, like the starting point for the PBC model, the model in Theorem 2 is also an adaptive Clifford circuit, but of a very restricted kind, having only $CZ$ gates and $X$ measurements. We can similarly commute all the $CZ$ gates out to the right beyond the final output measurements (and delete them there) to leave an adaptive sequence of Pauli measurements on an input state that's a product of the for 1-qubit states listed in Theorem 2 (thus more general than the $\ket{\pi/4}^{\otimes n}$ input for PBC).

However the Pauli measurements that arise in our model have only a very simple form because of two features:\\
(i) the only Clifford gates in the conjugations are $CZ$ gates and we have (for any lines $a$ and $b$)
\begin{equation}\label{czconj}
(CZ_{ab}) \,X_a\, (CZ_{ab}) = X_a\otimes Z_b
\end{equation}
(and similarly for $X$ on line $b$  as $CZ$ is symmetric). Also $CZ$ commutes with $Z$ on any line.\\
(ii) All intermediate $X$ measurements in our model arise from gate gadgets and the  measured line is never used again.  Thus for example, if an $X$ measurement is done on a line $a$, then by Equation (\ref{czconj}), a $CZ$ conjugation action can convert this into an $X_a\otimes Z_b$ Pauli measurement. But line $a$ is then never again used, so $X_a$ in the latter measurement will never be further expanded by $CZ_{ac}$ for any line $c$.\\
It follows from (i) and (ii) that the only Pauli measurements arising in our model are those having exactly one $X$ Pauli term, or exactly one $X$ and one $Z$ Pauli term, with $I$ on all other lines in both cases.\\[1cm]

\noindent
{\Large\bf Acknowledgements}\\
This work was done mostly while the authors RJ and SG were visiting the Simons Institute for the Theory of Computing, supported by NSF QLCI Grant No. 2016245, and SS was visiting independently.
We acknowledge the benefits of in-person collaboration and facilities that the Simons Institute provided.
SS acknowledges support from the Royal Society University Research Fellowship, and EPSRC Reliable and Robust Quantum Computing grant EP/W032635/1. We thank Yuki Takeuchi for comments on an earlier draft leading to an improvement in the proof of Theorem 1(a), and  further remarks on resources in Theorem 2.

\end{document}